\documentclass[11pt,a4paper]{article}

\usepackage{epsfig}
\usepackage{eucal}
\usepackage{amsfonts,amsmath}
\usepackage{mathrsfs,latexsym}
\usepackage{fancyhdr}
\usepackage{cite}

\newtheorem{rem}{Remark}[section]

\newtheorem{ex}{Example}[section]

\def\bbf{{\Bbb F}}

\def\bbc{{\Bbb C}}
\def\bbr{{\Bbb R}}
\def\bbz{{\Bbb Z}}

\def\tr{\mathrm{tr\,}}

\def\ad{\mathrm{ad\,}}
\def\Ad{\mbox{Ad}\,}

\def\mod{\text{mod\,}}

\def\rmi{\mathrm{i}}

\def\openone{\leavevmode\hbox{\small1\kern-3.3pt\normalsize1}}
\def\diag{\mbox{diag\,}}

\begin{document}
\title{New Reductions of a Matrix Generalized Heisenberg Ferromagnet
Equation}
\author{T. I. Valchev$^1$ and A. B. Yanovski$^2$\\
\small $^1$ Institute of Mathematics and Informatics,\\
\small Bulgarian Academy of Sciences, Acad. G. Bonchev Str., 1113 Sofia, Bulgaria\\
\small $^2$ Department of Mathematics \& Applied Mathematics,\\
\small University of Cape Town, Rondebosch 7700,
Cape Town, South Africa\\
\small E-mails: tiv@math.bas.bg, Alexandar.Ianovsky@uct.ac.za}
\date{}
\maketitle

\begin{abstract}
We present in this report $1+1$ dimensional nonlinear partial differential
equation integrable through inverse scattering transform. The integrable system under
consideration is a pseudo-Hermitian reduction of a matrix generalization of classical
$1+1$ dimensional Heisenberg ferromagnet equation. We derive recursion operators and
describe the integrable hierarchy related to that matrix equation.
\end{abstract}

integrable hierarchy\\


\section{Introduction}

The Heisenberg ferromagnet equation (HF)
\begin{equation}
\mathbf{S}_t=\mathbf{S}\times \mathbf{S}_{xx}  , \qquad
\mathbf{S}^2=1
\label{HFeq}\end{equation}
is one of classical equations integrable through inverse scattering transform \cite{blue-bible}.
Above, $\mathbf{S} = (S_1, S_2,S_3)$ is the spin vector of a one-dimensional ferromagnet and subscripts
mean partial derivatives with respect to space variable $x$ and time $t$, see \cite{borpop} for more details.
HF can be written as the compatibility condition $[L(\lambda), A(\lambda)] = 0$ of the Lax operators:
\begin{eqnarray*}
L(\lambda) & = & \rmi\partial_x-\lambda S ,\\
A(\lambda) & = & \rmi\partial_t + \frac{\rmi\lambda}{2}[S,S_x] + 2\lambda^2 S
\end{eqnarray*}
where $\lambda\in \bbc$ is spectral parameter, $\rmi=\sqrt{-1}$ and
\[S = \left(\begin{array}{cc}
S_3 & S_1 - \rmi S_2 \\ S_1 + \rmi S_2 & - S_3
\end{array}\right) .\]
In a series of papers \cite{gmv,side9,yan,yantih,yanvil}, the properties of the
pair of nonlinear evolution equations (NLEEs)
\begin{equation}
\begin{split}
\rmi u_t + u_{xx}+(\epsilon uu^*_x +  vv^*_x)u_x
+ (\epsilon uu^*_x + vv^*_x)_x u & = 0,\qquad \epsilon = \pm 1\\
\rmi v_t + v_{xx}+(\epsilon uu^*_x + vv^*_x)v_x
+ (\epsilon uu^*_x + vv^*_x)_x v & = 0
\end{split}
\label{ghf}
\end{equation}
and the auxiliary spectral problem associated with it have extensively been studied. Above, $*$
denotes complex conjugation and the complex-valued functions $u$ and $v$ are subject to the
condition $\epsilon|u|^2+|v|^2 = 1$. Obviously, we have two distinct systems here: one for
$\epsilon=+1$ (Hermitian reduction) and another for $\epsilon=-1$ (pseudo-Hermitian reduction).

Similarly to HF, (\ref{ghf}) can be written into the Lax form $[L(\lambda),A(\lambda)]=0$ for Lax
operators given by:
\begin{eqnarray*}
L(\lambda) & = & \rmi\partial_x - \lambda S, \qquad \lambda\in\bbc,\qquad
S = \left(\begin{array}{ccc}
	0 & u & v\\ 
	\epsilon u^* & 0 & 0 \\
	v^* & 0 & 0\end{array}\right),\\
A(\lambda) &=& \rmi\partial_t + \lambda A_1 + \lambda^2 A_2,
\qquad A_2  = \left(\begin{array}{ccc}
	- 1/3 & 0 & 0 \\ 0 & 2/3 - \epsilon|u|^2 & -\epsilon u^*v \\
	0 & - v^*u & 2/3 -|v|^2
\end{array}\right),\\
A_1 & = & \left(\begin{array}{ccc}
	0 & a & b\\ 
	\epsilon a^* & 0 & 0 \\
	b^* & 0 & 0\end{array}\right),\quad
\begin{array}{ccc} 
	a & = & - \rmi u_x - \rmi\left(\epsilon uu^*_x + vv^*_x\right)u\\
	b & = & - \rmi v_x - \rmi\left(\epsilon uu^*_x + vv^*_x\right)v
\end{array}.
\end{eqnarray*}
Thus, one can view (\ref{ghf}) as a formal $S$-integrable generalization of HF.

In the present report, we intend to consider a matrix version of (\ref{ghf}) and
discuss some of its basic properties. The matrix NLEE we aim to study still has a Lax
representation and its Lax pair is related to symmetric spaces of the type 
$\mathrm{SU}(m + n)/\mathrm{S}(\mathrm{U}(m)\times \mathrm{U}(n))$. We are going to
describe the integrable hierarchy of NLEEs using recursion operators. Our approach will
not use the notion of gauge equivalent NLEEs and gauge equivalent Lax pairs. 

The report is organized as follows. Next section introduces the main object of study in the
report --- pseudo-Hermitian reduction of a matrix HF equation and its Lax representation. In
section \ref{sec_hier}, we describe an integrable hierarchy of NLEEs associated with the matrix
HF equation in terms of recursion operators. Section \ref{concl} contains some further discussion
and final remarks.

\section{Matrix HF Type Equations}
\label{sec_mhf}

In this section, we shall introduce a new multicomponent NLEE generalizing the coupled system (\ref{ghf}).
Since our analysis will require certain knowledge of Lie algebras and Lie groups, we refer to
the classical monograph \cite{goto} for more detailed explanations. Let us start with a few remarks
on the notations we intend to use.

Assume $\bbf$ is either the field of real or complex numbers. We are going to denote the linear space
of all $m\times n$ matrices with entries in $\bbf$ by $M_{m,n}(\bbf)$. $\mathrm{SL}(n,\bbf)$ and $\mathrm{SU}(n)$
will stand for the special linear group over $\bbf$ and the unitary group of order $n$ respectively while
$\mathfrak{sl}(n,\bbf)$ and $\mathfrak{su}(n)$ will denote the corresponding Lie algebras.
When this does not lead to any ambiguity, we shall drop from the notation the field of scalars.

Next, we shall write $\left(X^T\right)_{ij}: = X_{ji}$, $X\in M_{m,n}(\bbf)$ and 
$\left(X^{\dag}\right)_{ij}: = X^*_{ji}$, $X\in M_{m,n}(\bbc)$ for the transposition and Hermitian conjugation
of a matrix. Also, we are going to use the notation $\openone_m$ for the unit matrix in $M_{m,m}(\bbf)$ and $Q_m\in M_{m,m}(\bbf)$ will denote a diagonal matrix with entries equal to $\pm 1$ (we do not specify the number of positive entries here).

It is well-known that $X \to - X^T$, $X\in\mathfrak{sl}(n)$ is an outer Lie algebra automorphisms
for $\mathfrak{sl}(n)$. An example of inner Lie algebra automorphism is given by the adjoint action
\[X\to \Ad_g X:= gXg^{-1}, \qquad g\in\mathrm{SL}(n), \qquad X\in\mathfrak{sl}(n)\]
of $\mathrm{SL}(n)$ on its Lie algebra. The derivation corresponding to the adjoint group action will
be denoted by $\ad_X(Y): = [X,Y]$, $X,Y\in \mathfrak{sl}(n)$.
 
Let us introduce the following Lax pair 
\begin{eqnarray}
L(\lambda) &:=& \rmi\partial_x - \lambda S,\qquad S := \left(\begin{array}{cc}
0 & \mathbf{u}^T \\ Q_n\mathbf{u}^*Q_m & 0
\end{array}\right),\label{lax1}\\
A(\lambda) &:=& \rmi\partial_t + \lambda A_1 + \lambda^2 A_2,\qquad
A_2 := \frac{2m}{m+n}\openone_{m+n} - S^{\, 2},\label{lax2}\\
A_1 & := & \left(\begin{array}{cc}
0 & \mathbf{a}^T \\ Q_n\mathbf{a}^*Q_m  & 0
\end{array}\right),\qquad \mathbf{a} := -\rmi(\mathbf{u}_x
+ \mathbf{u}Q_m\mathbf{u}^{\dag}_xQ_n\mathbf{u})\label{lax3}
\end{eqnarray}
where $\mathbf{u}:\bbr^2\to M_{n,m}(\bbc)$ is some smooth function.
We shall require that the matrix $\mathbf{u}(x,t)$ obeys the condition:
\begin{equation}
\mathbf{u}^T(x,t)Q_n\mathbf{u}^*(x,t)Q_m = \openone_m.
\label{constr}\end{equation}
Condition (\ref{constr}) can not be fulfilled for arbitrary matrices $Q_n$ and $Q_m$, e.g. if $Q_n = \openone_n$ and $Q_m = - \openone_m$ then (\ref{constr}) leads to contradiction. We shall assume that $Q_n$ and $Q_m$ are such that (\ref{constr}) leads to meaningful constraints for $\mathbf{u}(x,t)$.

Equation (\ref{constr}) represents orthonormality condition for the columns of \, $\mathbf{u}(x,t)$ viewed as vectors with respect to some pseudo-Hermitian form defined in $\bbc^n$ by the diagonal matrix $Q_n$. As a result, (\ref{constr}) can be satisfied for  $m \leq n$ only. However, when $\mathbf{u}(x,t)$ is a square matrix ($m = n$) then (\ref{constr}) gives rise to trivial flow. Indeed, we observe first that $\tr(Q_n\mathbf{u}^*Q_m\mathbf{u}^T) = \tr(\mathbf{u}^TQ_n\mathbf{u}^*Q_m) = m$. Next, condition (\ref{constr}) leads to the following relation: 
\begin{equation}
\left(Q_n\mathbf{u}^*(x,t)Q_m\mathbf{u}^T(x,t)\right)^2 = Q_n\mathbf{u}^*(x,t)Q_m\mathbf{u}^T(x,t).	
\label{vu_proj}\end{equation}
Therefore, when $m = n$ $Q_n\mathbf{u}^*(x,t)Q_m\mathbf{u}^T(x,t)$ has the maximal possible rank, i.e. it is invertible, and (\ref{vu_proj}) shows that it is simply the unit matrix. Thus, it is easily seen that $S^2 = \openone_{m+n}$ and $A_2$ becomes equal to zero. So in order to have a nontrivial construction we assume at that point that $m<n$
is fulfilled.

The compatibility condition $[L(\lambda),A(\lambda)] = 0$ of (\ref{lax1}), (\ref{lax2}) and
(\ref{lax3}) leads to the following matrix NLEE
\begin{equation}
\rmi \mathbf{u}_t + \mathbf{u}_{xx} + \left(\mathbf{u}Q_{m}\mathbf{u}^{\dag}_xQ_n\mathbf{u}\right)_x = 0.
\label{ghf_matr}
\end{equation}
Equation (\ref{ghf_matr}) turns into the matrix equation $(1)$ that appeared in \cite{gmv}
when setting $Q_m=\openone_m$ and $Q_n=\openone_n$.

Constraint (\ref{constr}) imposes a restriction on the spectrum of $S(x,t)$. Indeed, it is easy to
check that 
\begin{equation}
S^{\,3} = S,
\label{S_rel}\end{equation}
hence the eigenvalues of $S$ are $1$, $0$ and $-1$.

\begin{ex}
Let us consider the case when $m = 1$ and $n\geq 2$, i.e. $\mathbf{u}$ is a $n$-component vector function.
Without any loss of generality we can set $Q_1 = 1$ and assume that at least one diagonal entry of
$Q_n$ is $1$. Then (\ref{ghf_matr}) acquires the following vector form:
\begin{equation}
\rmi \mathbf{u}_t + \mathbf{u}_{xx} + \left(\mathbf{u}\mathbf{u}^{\dag}_xQ_n\mathbf{u}\right)_x	= 0	
\label{ghf_vec}\end{equation}
where $\mathbf{u}$ must satisfy 
\begin{equation}
\mathbf{u}^TQ_n\mathbf{u}^* = 1.
\label{constr_vec}\end{equation} 
Relation (\ref{constr_vec}) represents geometrically a sphere embedded in $\bbr^{2n}$ provided $Q_n = \openone_n$ and a hyperboloid in $\bbr^{2n}$ otherwise. Equation (\ref{ghf_vec}) and its anisotropic deformation were first considered by Golubchik and Sokolov \cite{golsok}.

In the vector case, the eigenvalues $\pm 1$ appear once in the spectrum of $S$ while $0$ has multiplicity $n-1$,
therefore one can pick up $\diag(1,0,\ldots,0,-1)$ as a canonical form of $S$. Evidently, for $n=2$ the vector
equation reduces to (\ref{ghf}).
\end{ex}

Let us now consider the case when $\mathbf{u}(x,t)$ is a rectangular matrix with $m>n$.
Now, we replace (\ref{constr}) with the following requirement:
\begin{equation}
Q_n\mathbf{u}^*Q_m\mathbf{u}^T = \openone_n
\label{constr2}
\end{equation}
and the second Lax operator (\ref{lax2}), (\ref{lax3}) with the following one:
\begin{eqnarray}
A(\lambda) &:=& \rmi\partial_t + \lambda A_1 + \lambda^2 A_2,\qquad
A_2 := \frac{2n}{m+n}\openone_{m+n} - S^{\, 2},\label{lax2b}\\
A_1 & := & \left(\begin{array}{cc}
0 & \mathbf{a}^T \\ Q_n\mathbf{a}^*Q_m  & 0
\end{array}\right),\qquad \mathbf{a} := \rmi(\mathbf{u}_x
+ \mathbf{u}Q_m\mathbf{u}^{\dag}_xQ_n\mathbf{u}).\label{lax3b}
\end{eqnarray}  
The compatibility condition of modified Lax pair (\ref{lax1}), (\ref{lax2b}) and (\ref{lax3b}) now gives
\begin{equation}
\rmi \mathbf{u}_t - \mathbf{u}_{xx} - \left(\mathbf{u}Q_{m}\mathbf{u}^{\dag}_xQ_n\mathbf{u}\right)_x = 0.
\label{ghf_matr2}
\end{equation}

Finally, consider the case when $m=n$. Conditions (\ref{constr}) and (\ref{constr2}) coincide for quadratic matrices and, as discussed earlier in text, and lead to trivial flows. This is why we need to impose another
(weaker) condition for $\mathbf{u}$, namely:
\begin{equation}
\mathbf{u}Q_m\mathbf{u}^{\dag}Q_n\mathbf{u} = \mathbf{u}.	
\label{constr3}\end{equation}

\begin{rem}
Evidently, equations (\ref{constr}) and (\ref{constr2}) give rise to (\ref{constr3}), i.e. they are special cases of it. In fact, for $\mathbf{u}(x,t)$ being a $n$-vector ($m=1$) constraint (\ref{constr3}) is equivalent to (\ref{constr_vec}). On the other hand, if $\mathbf{u}(x,t)$ is an invertible square matrix then (\ref{constr3}) is reduced to (\ref{constr}) (or equivalently to (\ref{constr2})). This is why we shall be interested in 
the case when $\mathbf{u}(x,t)$ is not invertible.
\label{rem3}\end{rem} 

It is easy to see that equation (\ref{S_rel}) holds if and only if $\mathbf{u}$ satisfies (\ref{constr3}), i.e. (\ref{constr3}) is the weakest condition leading to (\ref{S_rel}). 
Thus, $S(x,t)$ can be diagonalized and its spectrum consists of $0$, $-1$ and $1$.

Condition (\ref{constr3}) requires certain modification in the second Lax operator --- we have to use now the following operator:
\begin{eqnarray}
A(\lambda) &:=& \rmi\partial_t + \lambda A_1 + \lambda^2 A_2,\qquad
A_2 := \frac{2r}{m+n}\openone_{m+n} - S^{\, 2},\label{lax2c}\\
A_1 & := & \left(\begin{array}{cc}
0 & \mathbf{a}^T \\ Q_n\mathbf{a}^*Q_m  & 0
\end{array}\right),\qquad \mathbf{a} := \rmi\left[ \mathbf{u}(Q_m\mathbf{u}^{\dag}Q_n\mathbf{u})_x - (\mathbf{u}Q_m\mathbf{u}^{\dag}Q_n)_x\mathbf{u} \right]\label{lax3c}
\end{eqnarray} 
instead of (\ref{lax2}), (\ref{lax3}) (or (\ref{lax2b}), (\ref{lax3b})). The number $r: = \tr(\mathbf{u}^TQ_n\mathbf{u}^*Q_m)$ is assumed to be not greater than $\min(m,n)$, for $m\neq n$ and strictly less than $m$ for $m=n$, see Remark \ref{rem3}. The compatibility condition of (\ref{lax1}) and (\ref{lax2c}), (\ref{lax3c}) now leads to
\begin{equation}
\rmi\mathbf{u}_t + \left[(\mathbf{u}Q_m\mathbf{u}^{\dag}Q_n)_x\mathbf{u} - \mathbf{u}(Q_m\mathbf{u}^{\dag}Q_n\mathbf{u})_x\right]_x = 0.
\label{ghf_matr3}\end{equation}
Obviously, (\ref{ghf_matr3}) includes (\ref{ghf_matr}) and (\ref{ghf_matr2}) as particular cases.

The form of the matrix coefficients in (\ref{lax1}), (\ref{lax2}), (\ref{lax3}), (\ref{lax2b}), (\ref{lax2c})
and (\ref{lax3c}) implies that the Lax operators are subject to the following symmetry conditions:
\begin{eqnarray}
HL(-\lambda)H &=&  L(\lambda),\qquad HA(-\lambda)H = A(\lambda),\label{red1}\\
Q_{m+n}L^{\dag}(\lambda^*)Q_{m+n}  &=& -\tilde{L}(\lambda) ,\qquad
Q_{m+n}A^{\dag}(\lambda)Q_{m+n} = -\tilde{A}(\lambda)\label{red2}
\end{eqnarray}
where $H=\diag(-\openone_m,\openone_n)$, $Q_{m+n} = \diag(Q_m,Q_n) $ and
$\tilde{L}\psi: = \rmi\partial_x\psi + \lambda\psi S$. The ajoint action of $H$
in $\mathfrak{sl}(m+n,\bbc)$ is involutive, hence it defines a $\bbz_2$-grading
of the Lie algebra in the following way: 
\[\begin{split}
\mathfrak{sl}(m+n) &= \mathfrak{sl}^0(m+n) \oplus \mathfrak{sl}^1(m+n),\\
\mathfrak{sl}^{\sigma}(m+n) &:= \{X\in\mathfrak{sl}(m+n): \; \Ad_HX = (-1)^{\sigma}X\},
\qquad \sigma = 0,1.
\end{split}\]
The eigenspace $\mathfrak{sl}^0(m+n)$ consists of block diagonal matrices with $m\times m$ and
$n\times n$ blocks on its principal diagonal, e.g. the matrix coefficient $A_2$, while
$\mathfrak{sl}^1(m+n)$ is spanned by matrices of a form like $S$ and $A_1$. Let us also remark
that the relation $S=-HSH$ shows that $S$ is a semisimple matrix, hence it is diagonalizable.

The matrix $H$ is deeply related to the Cartan involution underlying the definition of the symmetric space
$\mathrm{SU}(m + n)/\mathrm{S}(\mathrm{U}(m)\times \mathrm{U}(n))$. This is why we say that the Lax pair
(\ref{lax1})--(\ref{lax3}) is related to the symmetric space 
$\mathrm{SU}(m + n)/\mathrm{S}(\mathrm{U}(m)\times \mathrm{U}(n))$ following the convention proposed by
Fordy and Kulish \cite{ForKu}.

Using the adjoint action of $Q_{m+n}$ as appeared in (\ref{red2}), one can introduce a complex conjugation
$\mathcal{I}$ in $\mathfrak{sl}(m+n)$ setting: 
\[\mathcal{I}(X) := - Q_{m+n}X^{\dag}Q_{m+n},\qquad X\in\mathfrak{sl}(m+n).\]
This complex conjugation defines the compact real form $\mathfrak{su}(m+n)$ of  $\mathfrak{sl}(m+n)$
in case $Q_{m+n}= \openone_{m+n}$ or the real form $\mathfrak{su}(l,m+n-l)$ in case the matrix $Q_{m+n}$
has $l$ diagonal entries equal to $1$. In order to treat both cases simultaneously we shall refer
to condition (\ref{red2}) as a pseudo-Hermitian reduction when $Q_{m+n}\neq \openone_{m+n}$ and a Hermitian
one when $Q_{m+n}= \openone_{m+n}$. 

Let us consider the linear problem
\begin{equation}
L(\lambda)\psi(x,t,\lambda) = 0
\label{aux_pr}\end{equation}
and denote the set of its fundamental solutions by $\mathcal{F}$. Then (\ref{red1}) and (\ref{red2}) 
could be considered consequences of a Mikhailov-type reduction imposed on the linear spectral problem,
see \cite{mik,yantih}. Indeed, consider the following maps:
\begin{eqnarray}
\psi_0(\lambda) &\to& G_1\psi_0(\lambda)= H\psi_0(-\lambda)H,\label{psi_red1}\\
\psi_0(\lambda) &\to&  G_2\psi_0(\lambda)=Q_{m+n}\left[\psi^{\dag}_0(\lambda^*)\right]^{-1}Q_{m+n}
\label{psi_red2}\end{eqnarray}
and assume that $\mathcal{F}$ is invariant under $G_1$ and $G_2$. Since $G_1G_2=G_2G_1$, $G_1^2=G_2^2={\rm id}$,
(\ref{psi_red1}) and (\ref{psi_red2}) define an action of the Mikhailov reduction group $\bbz_2\times\bbz_2$ for
the spectral problem (\ref{aux_pr}). As it is easily seen, this leads to (\ref{red1}) and (\ref{red2}) respectively.

\section{Integrable Hierarchy and Recursion Operators}
\label{sec_hier}

In this section, we shall describe the hierarchy of matrix integrable NLEEs associated with matrix
equation (\ref{ghf_matr}). In doing this, we shall follow ideas and methods discussed in \cite{book,side9}. 

Let us consider the general flow Lax pair:
\begin{equation}
\begin{split}
L(\lambda) :=& \rmi\partial_x - \lambda S,\\
A(\lambda) :=& \rmi\partial_t + \sum^N_{j=1}\lambda^j A_j,\qquad N\geq 2
\label{ghf_lax_gen}
\end{split}
\end{equation}
which is subject to reduction conditions (\ref{red1}) and (\ref{red2}). The matrix $S(x,t)$ has
the same form as in (\ref{lax1}) and it is assumed to obey (\ref{S_rel}), i.e. constraint
(\ref{constr3}) holds for $\mathbf{u}(x,t)$. Then the zero curvature condition $[L(\lambda), A(\lambda)] = 0$ of the Lax pair (\ref{ghf_lax_gen}) gives
rise to the following set of recurrence relations:
\begin{eqnarray}
&&	[S, A_N] = 0 , \label{rec_np1}\\
&& \ldots \nonumber\\
&& \rmi\partial_xA_k - [S,A_{k-1}] = 0,
\qquad k=2,\ldots, N ,\label{rec_k}\\
&& \ldots \nonumber\\
&& \partial_xA_1 + \partial_t S = 0.\label{rec_1}
\end{eqnarray}
Each solution to these equations leads to a member of the integrable hierarchy of NLEEs. The
analysis of (\ref{rec_np1})--(\ref{rec_1}) resembles very much the one we have in the case of
$2$-component system (\ref{ghf}). Following \cite{side9,yantih},we introduce the splitting
\begin{equation}
A_j = A^{\rm a}_j + A^{\rm d}_j\ , \qquad j=1,\ldots,N
\label{a_split}\end{equation}
of the matrix coefficients of $A$ into a $S$-commuting term $A^{\rm d}_j$ and some remainder
$A^{\rm a}_j$. We discussed in the previous section that $S(x,t)$ is a diagonalizable matrix. Hence
$\ad_S$ is diagonalizable too with eigenvalues $0,\pm 1, \pm 2$. The above splitting simply says that
$ A^{\rm d}_j$ and $ A^{\rm a}_j$ belong to the zero eigenspace of $\ad_S$ and the direct sum of all the nonzero eigenspaces respectively. Consequently, the above splitting is unique and the operator $\ad_S^{-1}$
is properly defined on $ A^{\rm a}_j$. Moreover, considering the minimal polynomial of $\ad_S$
on the direct sum of all nonzero eigenspaces, one gets $(\ad_S^2-1)(\ad_S^2-4)=0$. Therefore, we have
\[\ad^{-1}_{S} = \frac{1}{4}\left(5\ad_{S} - \ad^3_{S}\right).\]

Further, symmetry conditions (\ref{red1}) and (\ref{red2}) require that we have 
\[ A_j\in \mathfrak{sl}^{\sigma}(m+n,\bbc)\cap \mathfrak{su}(l,m+n-l), \qquad j \equiv \sigma \quad(\mod 2), \qquad \sigma = 0, 1 \]
for the coefficients of the second Lax operator, see the comments at the bottom of page $5$. For our purposes it will be enough to take
\begin{equation}
A^{\rm d}_{j} = \begin{cases}
a_{j} S_1,& j \equiv 0 \quad(\mod 2)\\ a_{j} S,& j\equiv 1 \quad (\mod 2) 
\end{cases}
\label{a_j_diag}\end{equation}
where
\[S_1 := S^{\,2} - \frac{2r}{m+n}\openone_{m+n}\]
and $a_{j}$, $j=1,\ldots,N$ are some scalar functions to be determined in such a way that the recurrence
relations are satisfied.

Equation (\ref{rec_np1}) means that $A^{\rm a}_N = 0$ and in accordance with our previous assumption,
for the highest degree coefficient we may take:
\[A_{N} = \begin{cases} 
c_NS ,& N \equiv 1 \quad (\mod 2) \\ c_N S_1 ,& N \equiv 0 \quad (\mod 2)
\end{cases},\qquad c_N\in\bbr.\]

Let us consider now recurrence relation (\ref{rec_k}). After substituting (\ref{a_split})
into (\ref{rec_k}) and taking into account that 
\[(S_x)^{\rm d} =  (S_{1,x})^{\rm d} = 0, \]
we derive the following two relations
\begin{equation}
\left(\partial_xA^{\rm a}_k\right)^{\rm d} = - \begin{cases} 
\partial_xa_k S ,& k \equiv 1 \quad (\mod 2) \\ \partial_xa_k S_1 ,& k \equiv 0 \quad (\mod 2) \end{cases}\; ,\qquad k=2,\ldots,N
\label{rec_k_d}
\end{equation}
for the $S$-commuting terms and
\begin{equation}
\left[S, A^{\rm a}_k\right] - \rmi\left(\partial_xA^{\rm a}_k\right)^{\rm a} =
\begin{cases} 
\rmi a_k S_x ,& k \equiv 1 \quad (\mod 2) \\ \rmi a_k S_{1,x} ,& k \equiv 0 \quad (\mod 2) \end{cases}
\label{rec_k_a}
\end{equation}
for the terms which do not commute with $S$. In order to solve (\ref{rec_k_d}), we also make use of the normalization conditions:
\[\tr S^2 = 2r\, ,\qquad \tr S^2_1 = \frac{2r(m+n-2r)}{m + n}\, .	\]
As a result, we obtain the following expression for $a_k$
\begin{equation}
a_k = c_k - \begin{cases} 
\frac{1}{2r}\,\partial^{-1}_x\tr\left[S(\partial_xA_k)^{\rm d}\right] ,& k \equiv 1 \quad (\mod 2) \\ \frac{m+n}{2r(m+n-2r)}\,\partial^{-1}_x\tr\left[S_1(\partial_xA_k)^{\rm d}\right] ,& k \equiv 0 \quad (\mod 2) \end{cases}	
\label{a_k}\end{equation}
where the symbol $\partial^{-1}_x$ stands for any right inverse of the operator of partial differentiation in variable $x$ and $c_k\in\bbr$ is an integration constant. After substituting (\ref{a_k}) into (\ref{rec_k_a}), we obtain 
\[A^{\rm a}_{k-1} = \begin{cases}
\Lambda A^{\rm a}_{k} + \rmi c_k\ad^{-1}_{S}S_{1,x}\, ,&
 k\equiv 0 \quad (\mod 2)\\
\Lambda A^{\rm a}_{k} + \rmi c_k\ad^{-1}_{S}S_{x}\, ,&
k\equiv 1 \quad (\mod 2) 
\end{cases}\]
where  
\begin{eqnarray*}
\Lambda &:=& \rmi\ad^{-1}_{S}\left\{ [\partial_x(.)]^{\rm a} - \frac{S_{x}}{2r}
\partial^{-1}_x\tr\left[S(\partial_x(.))^{\rm d}\right] - \frac{(m+n)S_{1,x}}{2r(m+n-2r)}
\partial^{-1}_x\tr\left[S_1(\partial_x(.))^{\rm d}\right]\right\}.
\end{eqnarray*}
The $\Lambda$ operator as defined above acts on the $S$-non commuting part of $A_j$ only. However,
one can extend the action of $\Lambda$ on the $S$-commuting part as well by requiring 
\[\Lambda S :=  \rmi \ad^{-1}_{S} S_{x}\, ,\qquad
\Lambda S_{1} :=  \rmi \ad^{-1}_{S} S_{1,x}\,.\]
Then an arbitrary member of the integrable hierarchy we consider can be written down as follows:
\begin{equation}
\rmi\ad^{-1}_{S} S_{t}	+ \sum_{k}c_{2k}\Lambda^{2k}S_1
+ \sum_{k}c_{2k-1}\Lambda^{2k-1} S = 0.
\label{int_hier}\end{equation}
The operator $\Lambda^2$ is called recursion operator of the above hierarchy of NLEEs. It is easy
to check that (\ref{int_hier}) gives (\ref{ghf_matr}) after setting $N=2$, $c_2 = -1$ and $c_1 = 0$.
Thus, it is the simplest representative of the family (\ref{int_hier}).

\section{Conclusion}
\label{concl}

In the present report, we have considered new multicomponent NLEE of HF type, see (\ref{ghf_matr}),
(\ref{ghf_matr2}) and (\ref{ghf_matr3}). That NLEE is $S$-integrable with a Lax pair associated with
the symmetric space $\mathrm{SU}(m + n)/\mathrm{S}(\mathrm{U}(m)\times \mathrm{U}(n))$, see
(\ref{lax1})--(\ref{lax3}), (\ref{lax2b}), (\ref{lax3b}), (\ref{lax2c}) and (\ref{lax3c}). Thus,
it is a natural generalization of the coupled system (\ref{ghf}) already studied. Moreover,
we have derived recursion operators that allowed us to describe an integrable hierarchy related to
(\ref{ghf_matr}), see (\ref{int_hier}). We would like to stress here on an important difference between
the integrable hierarchy for (\ref{ghf}) as obtained in \cite{yantih} and that one we have derived in previous section. Unlike the integrable hierarchy for the coupled system, (\ref{int_hier}) is {\bf not} the most general one. Indeed, one may observe that $\ker\ad_S$ is not spanned by $S$ and $S_1$ only. This means that when we have picked up $A^{\rm d}_j$ in the form (\ref{a_j_diag}) we have neglected some $S$-commuting terms which contribute to the $\Lambda$-operators, therefore contribute to the form of the integrable hierarchy.

As discussed in \cite{side9}, the recursion operator of a S-integrable NLEE could also be derived through a method which is a modification of the method used in the previous section. For its implementation in the case of scattering operator (\ref{lax1}), one starts from a Lax representation for two adjacent equations in the hierarchy with evolution parameters $t$ and $\tau$ respectively, see \cite{GKS}. Recursion operator $\mathcal{R}$ is then viewed as the mapping
\[S_{\tau} = \mathcal{R} S_{t}\, .\]
Then one can prove that $\mathcal{R}$ and $\Lambda^2$ as derived in main text are interrelated through
\[\mathcal{R} = \ad_{S}\Lambda^2\,\ad^{-1}_{S}\, .\]
Therefore they have essentially the same properties and are equivalent. 

We have not discussed in detail some important issues like the spectral properties of the scattering
operator. As it is well-known, the spectrum of $L$ depends on the asymptotic behavior of potential
function $\mathbf{u}$ and the reductions imposed on the Lax pair. In the simplest case of constant boundary conditions, one can prove that the continuous spectrum of (\ref{lax1}) coincides with the real line
in $\bbc$, see \cite{yantih}. On the other hand, the discrete eigenvalues of $L$ will be located symmetrically with respect to the real and imaginary lines in $\bbc$ due to reductions (\ref{red1}) and (\ref{red2}).
 
Another important issue concerns the solutions of (\ref{ghf_matr}). Similarly to the $2$-component case, see
\cite{yantih2}, we can distinguish between solutions of soliton type and solutions of quasi-rational type.
The latter correspond to a degenerate spectrum of the scattering operator \cite{tiv_ntades15}. We intend to address all these issues elsewhere.

\section*{Acknowledgments}

The work has been supported by the NRF incentive grant of South Africa and grant DN 02-5 of Bulgarian
Fund "Scientific Research".

\end{document}